# Hydrodynamic charge transport in GaAs/AlGaAs ultrahigh-mobility two-dimensional electron gas

Xinghao Wang, Peizhe Jia, and Rui-Rui Du[*]

International Center for Quantum Materials, School of Physics, Peking University, Beijing 100871, China

L. N. Pfeiffer, K. W. Baldwin, and K. W. West

Department of Electrical Engineering, Princeton University, Princeton, NJ 08544, USA

## Abstract

Viscous fluid in an ultrahigh-mobility two-dimensional electron gas (2DEG) in GaAs/AlGaAs quantum wells is systematically studied through measurements of negative magnetoresistance (NMR) and photoresistance under microwave radiation, and the data are analyzed according to recent theorical work by *e.g.*, Alekseev, Physical Review Letters **117**,166601 (2016). Size-dependent and temperature dependent NMR are found to conform to the theoretical predictions. In particular, transport of 2DEG with relatively weak Coulomb interaction (interparticle interaction parameter $r_s < 1$) manifests a crossover between viscous liquid and viscous gas. The size dependence of microwave induced resistance oscillations and that of the '2nd harmonic' peak indicate that 2DEG in a moderate magnetic field should be regarded as viscous fluid as well. Our results suggest that the hydrodynamic effects must be considered in order to understand semiclassical electronic transport in a clean 2DEG.

[*] rrd@pku.edu.cn

*Introduction.* -Viscosity in electron flow results from shear stress among electrons and the corresponding charge transport is within the hydrodynamic regime. More than half a century ago, R. N. Gurzhi and coauthors predicted that viscous Poiseuille flow of electrons exist in 3D metals if $l_{ee} \ll d^2/l_{ee} \ll l_0$ [1], where $d$ is the sample thickness, and $l_{ee}$, $l_0$ is the mean free path (MFP) for quasiparticle (electron) collisions and bulk collisions respectively. The former MFP conserves quasiparticle momentum while the latter, including MFP for scatterings with impurities and phonons, breaks quasi-momentum conservation. Under this condition, sheer stress among electrons may result in hydrodynamic transport with many intriguing effects to be explored. Although such phenomenon was not observed in 3D metal due to a short $l_0$ induced by impurities and defects, clean 2DEG is proven to be a promising platform for studying hydrodynamics in electron systems [2–13]. Besides fundamental interest, hydrodynamic property of 2DEG has important ramifications on the performance of electronic devices such as nano-transistors or quantum point contacts.

Hydrodynamic charge transport in high-mobility 2D electron systems has been discussed theoretically for decades [2–9], but it does not catch great attention until recently, as evidence of viscous liquid has been discovered in 2D metal PdCoO$_2$ [10], graphene[11, 12] and GaAs/AlGaAs quantum well (QW)[13]. Graphene draws much interest in this field because its electron is scattered weakly by acoustic phonons so that viscosity of electrons is even observable at high temperatures [11, 12]. As the cleanest electron system available so far, ultrahigh-mobility GaAs/AlGaAs QW should be a natural material manifesting rich hydrodynamic properties through its prominent NMR [14–22] and exotic photoresistance (PR) induced by microwave radiation[14, 16, 17]. However, these phenomena were not understood as a result of viscosity of electron fluid in the early prints [14-16], rather, *e.g.*, as that of an interplay between smooth long-range disorder and rare strong scatterers in high-mobility samples [23], until theories of viscous flow in a moderate magnetic field (when Landau levels are not formed) in 2DEG were proposed by *e.g.*, Alekseev, and others [24–29].

For a sample with width $W$, in hydrodynamic regime where $l_{ee} \ll W^2/l_{ee} \ll l_0$, scattering among quasiparticles dominates in the diffusive transport. This gives rise to a charged Poiseuille flow with enhanced zero-field resistivity $\rho_{xx}(0) \propto W^{-2}$ and decreasing $\rho_{xx}(B)$ with a finite magnetic field since viscosity of electrons is greatly weakened by cyclotron motion[24]. Moreover, giant PR peak at microwave (MW) frequency $\omega = 2\omega_c$ ('2nd harmonic' for short) observed in [14, 16], where $\omega_c = eB/m^*$ is cyclotron frequency with effective electron mass $m^*$, can be explained qualitatively by the theory of transverse magnetosonic waves caused by shear stress in Fermi liquid[26]. More recently, hydrodynamics in 2DEG has been applied to microwave-induced resistance oscillations (MIRO)[30–32], and significant and quantitative predictions were proposed in [29]. The present experiments, specifically using ultrahigh-mobility GaAs/AlGaAs QW samples, are partly promoted by these theoretical works.

In this Letter, hydrodynamics of 2DEG in modulation-doped GaAs/AlGaAs QW is systematically studied in low-temperature transport experiments performed in a $^3$He refrigerator with three ultrahigh-mobility wafers (with a nominal mobility $\mu$ above $2 * 10^7 cm^2/Vs$ at 0.3 K), carrier density $n$ of which ranges from $2.0 * 10^{11}\ cm^{-2}$ to $4.2 * 10^{11}\ cm^{-2}$. Each sample consists of five sections of Hall bars with different sample widths ($W = 400, 200, 100, 50, 25\ \mu m$) and the same length-to-width ratio $L/W = 3$, defined by photolithography and wet etching. Electrical contacts were made by In/Sn alloy. Sample A and B are respectively from wafer A and B, and both Sample C1 and C2 are from wafer C prepared with different illumination conditions. Sample A, B and C2 are illuminated by red light-emitting diode at 2 K and Sample C1 remains in a dark environment during entire measurements. For details of each wafer and the structure of Hall bars, please refer to Table I and inset (a) of Fig.1. Transport data were collected by standard lock-in technique with an excitation current 1 μA and frequency 17 Hz. In all the four samples, huge NMR and sharp '2nd harmonic' peaks are observed for narrow Hall bars. Additionally, amplitude of MIRO indeed demonstrates the $W$-dependence conforming to the predictions by [29]. Altogether, the experimental results affirmatively and consistently support the viscous liquid theory of 2DEG[24–29].

TABLE I. Values of the parameters of all the four samples at 0.3 K. First three rows provide information of the three wafers. x is the mole fraction of Al in $Al_xGa_{1-x}$ As barriers. $w$ is well width and $d_s$ is spacer distance. $\mu$ represents sample mobility without the contribution from viscosity. $\tau_2$ is viscosity related relaxation time extracted from data of the $25\mu m$ Hall bar. $v_F^\eta$ is obtained through the inset of Fig.1(b). Fermi velocity $v_F$ is calculated from carrier density.

| At 0.3K | Sample A | Sample B | Sample C1 | Sample C2 |
|---|---|---|---|---|
| x | 0.294 | 0.158-0.24 * | 0.16-0.22-0.32 * | |
| w/nm | 30 | 29 | 26 | |
| $d_s$/nm | 81 | 82 | 81 | |
| n/$10^{11}$cm$^{-2}$ | 2.01 | 2.62 | 3.64 | 4.20 |
| $r_s$ | 1.24 | 1.09 | 0.92 | 0.86 |
| $\mu$/$10^6$cm$^2$V$^{-1}$s$^{-1}$** | 41.2 | 22.3 | 39.9 | 42.5 |
| $\tau_2$/$10^{-11}$s | 4.13 | 5.20 | 5.78 | 5.07 |
| $v_F^\eta$/$10^5$ms$^{-1}$ | 1.31 | 1.07 | 1.09 | 1.28 |
| $v_F$/$10^5$ms$^{-1}$ | 1.93 | 2.22 | 2.61 | 2.80 |

* Graded $Al_xGa_{1-x}As$ barrier

**Referring to Fig. 1 inset (b), $\rho_{xx}(0)$ without the contribution of viscosity is extrapolated from the data to $1/W = 0$. The respective $\mu$ is then calculated by $\mu = 1/ne\rho_{xx}(0)|_{W\to\infty}$.

*Negative magnetoresistance.* -As an example, Fig.1 demonstrates the key feature of NMR for different sample widths W. The NMR consists of two distinct parts: a narrow, sensitively sample-size-dependent peak (NNMR) dominating within ±100 G and a broad 'bell-shaped' NMR (BNMR) which is less dependent of $W$. These two coexisting parts were also reported in previous experiments[14–20, 22] and can be discriminated from each other, for BNMR is easily suppressed by adding a moderate in-plane magnetic field ($B_\parallel < 10\ kG$) or warming up to several kelvins[18], as clearly shown in Fig. 2(a) and (b). However, these early reports suggested that NNMR is either caused by weak localization or by oval defects in GaAs/AlGaAs QWs. In hindsight, this was partly because their samples were too wide to form a distinct NNMR, or the analysis confused NNMR with BNMR when only one kind of NMR dominates. Recently, NMR in high density 2DEG of GaAs/AlGaAs QW has been reported and explained within the framework of hydrodynamics[13, 33, 34], but BNMR was not observed. Later in this paper, NNMR is shown to be a direct result of viscoelastic dynamics of 2D electron fluid and BNMR, may be also related to hydrodynamical effect, seems to be more complex and obscure.

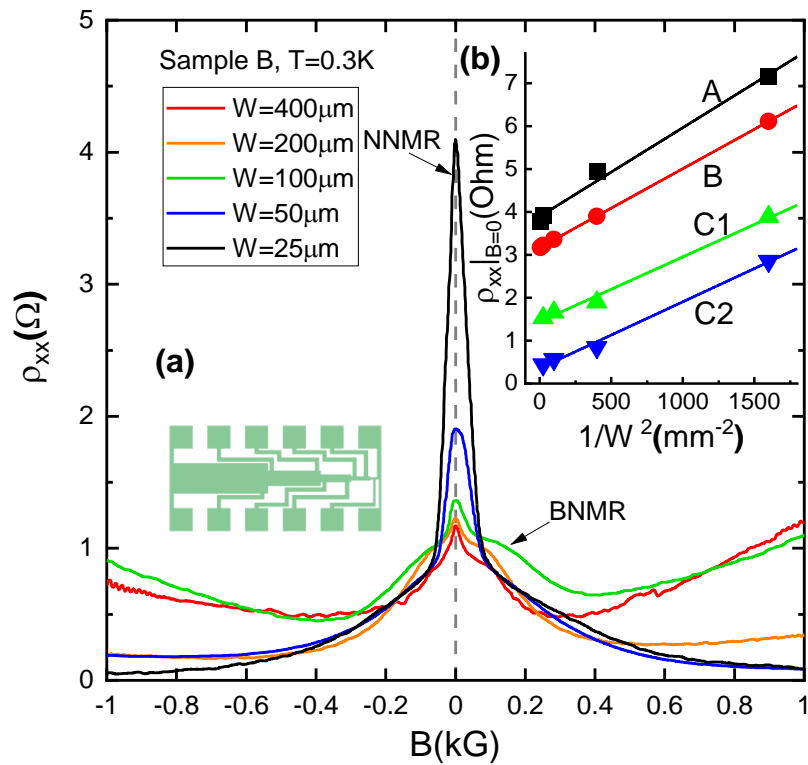

FIG. 1. NMR of Sample B as an example with Hall bar width ranges from 400 μm to 25 μm at 0.3 K. BNMR and NNMR can be distinguished in all five traces. Inset (a) shows configuration of the Hall bar in our experiment. Inset (b) demonstrates linear relation between zero-field resistivity and $1/W^2$ in all the four samples. For clarity, respective data for sample A, B, C1 is consecutively shifted upward by 1 Ω.

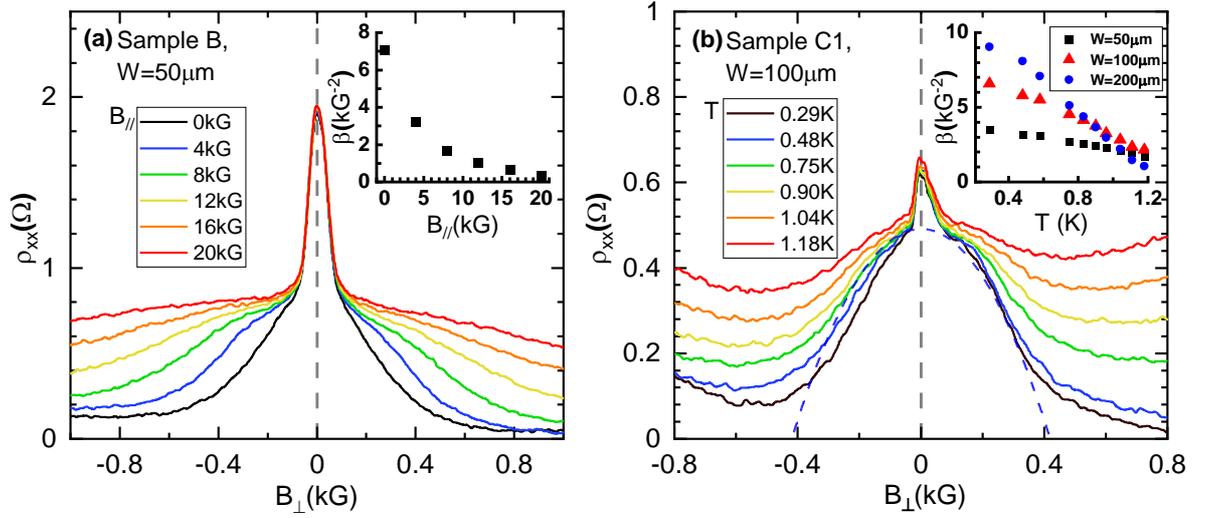

FIG. 2. (a) NNMR is not affected by in-plane magnetic field, but BNMR is. The data is measured from 50μm Hall bar of Sample B at 0.3 K. The inset shows the relation between $\beta$ from Eq. (4) and the in-plane field. (b) Higher temperature suppresses BNMR more quickly than NNMR. The data is measured from 100 $\mu m$ Hall bar of Sample C1. As an example, the blue dashed curve is a fit to the data at 0.48K. The inset demonstrates temperature dependence of $\beta$ with three different widths in Sample C1.

From theoretical perspectives, Poiseuille flow of electrons in Drude model results in an effective relaxation time $\tau^* = W^2/12\eta$, where $\eta = (v_F^\eta)^2 \tau_2/4$ represents viscosity of electron fluid[24]. $v_F^\eta$ is the Fermi velocity corrected by viscosity $\eta$[26]. $\tau_2$ is the relaxation time of the second moment of the electron distribution function, including contributions from the quasiparticle-quasiparticle (electron-electron) collisions $\tau_{2,ee}$ and the temperature-independent part $\tau_{2,0}$ determined by electron-disorder scattering. It has to be mentioned that $\tau_2$ satisfies reciprocal rule $1/\tau_2 = 1/\tau_{2,ee} + 1/\tau_{2,0}$. With a finite magnetic field, viscosity is attenuated due to loss of shear stress and magnetoresistivity is represented approximately as [24]

$$\rho_{xx}(B) = \frac{m^*}{e^2 n}\left(\frac{1}{\tau_0} + \frac{1}{\tau^*(1+(2\omega_c \tau_2)^2)}\right), \qquad (1)$$

where $m^* = 0.067 m_e$ is the effective mass of electrons in GaAs and $m_e$ is the mass of free electrons. Inset (b) of Fig.1 shows that zero-field resistivity $\rho_{xx}(0)$ is linear with $1/W^2$ in all the four samples, which is in correspondence with Eq.(1). This result indicates that NNMR matches the viscosity related NMR described in [24]. However, $v_F^\eta$ fitted in this method is smaller than Fermi velocity (Table. I), which quantitatively contradicts with the viscoelastic theory where viscosity and Fermi velocity are both enlarged in a strongly interacting nonideal Fermi liquid[26]. Since the boundary scattering may become important in small W, this inconsistency could be explained if we consider further correction $\tau^* = W(W + 6l_s)/12\eta$, where $l_s$ is boundary slip length[22] and the corresponding $v_F^\eta$ could be larger. This issue remains to be elucidated by future works.

Further evidence for viscosity induced NNMR is provided in Fig. 3 showing temperature dependence of $1/\tau_2$ in wafer B and C. Paper [28] suggested that electron systems in the two hydrodynamic regimes (liquid or gas) should crossover depending on the strength of Coulomb interactions. For low density 2DEG, interparticle interaction parameter $r_s = 1/(\sqrt{\pi n} a_B)$ is on the order of 1 ($a_B = 10.1 nm$, is Bohr radius in GaAs), and electron-electron scattering relaxation time [8]

$$\frac{1}{\tau_{2,ee}} \propto \frac{T^2}{\ln^2(\epsilon_F/k_B T)}, \tag{2}$$

where $T$ is electron temperature, $k_B$ refers to Boltzmann constant, and $\epsilon_F$ is Fermi energy of 2DEG. This regime is called viscous liquid for the strong interaction between particles. However, for high density 2DEG with $r_s \ll 1$, which is described as viscous gas, $\tau_{2,ee}$ can be expressed with a different logarithmic factor [28]:

$$\frac{1}{\tau_{2,ee}} = \frac{8\pi}{3\hbar} \frac{(k_B T)^2 r_s^2}{\epsilon_F} \ln\left(\frac{1}{r_s + T/\epsilon_F}\right), \tag{3}$$

where $\hbar$ represents Planck constant divided by $2\pi$. Despite the fact that density of ordinary ultrahigh-quality 2DEG in GaAs QW lies in the regime of viscous liquid and thus $\tau_{2,ee}$ should obey Eq. (2), viscous-gas-like temperature dependence of $\tau_2$ of Sample C1 with $r_s = 0.92$ and C2 with $r_s = 0.86$ seems to indicate electrons in it are within a crossover phase between the viscous liquid and the viscous gas described by Eq. (2) and (3). Relaxation time $\tau_2$ is extracted from full width half maximum of NNMR (Fig. 3(a)). In Fig.3(c), Sample B possesses a stronger Coulomb interaction ($r_s = 1.09$) and its $\tau_{2,ee}$ is fitted quite well with Eq. (2), confirming the hydrodynamic nature of NNMR. For wafer C (Fig. 3(d) and (e)), $\tau_{2,ee}$ is fitted precisely with Eq. (3) as long as $1 - r_s \gg T/\epsilon_F$ (so that temperature dependence in logarithmic factor of Eq. (3) can be ignored) but much off with Eq. (2). Notice that fitting $\tau_{2,ee}$ with Eq. (3) actually uses no adjustable parameter, which means at low temperatures Eq. (3) accurately describes the interparticle relaxation time in wafer C. This result clearly confirms the viscoelastic dynamics theory in [28] but, a caveat should be noted here. Despite its viscous-gas like temperature dependence of $\tau_2$, Sample C1 and C2 cannot be characterized into viscous gas because strong '2nd harmonic' peaks are discovered in them (to be presented in Fig.4(b)), which is a typical feature of viscous liquid [28]. Our results present an interesting case for further studies into the crossover region where existing theory does not fully address.

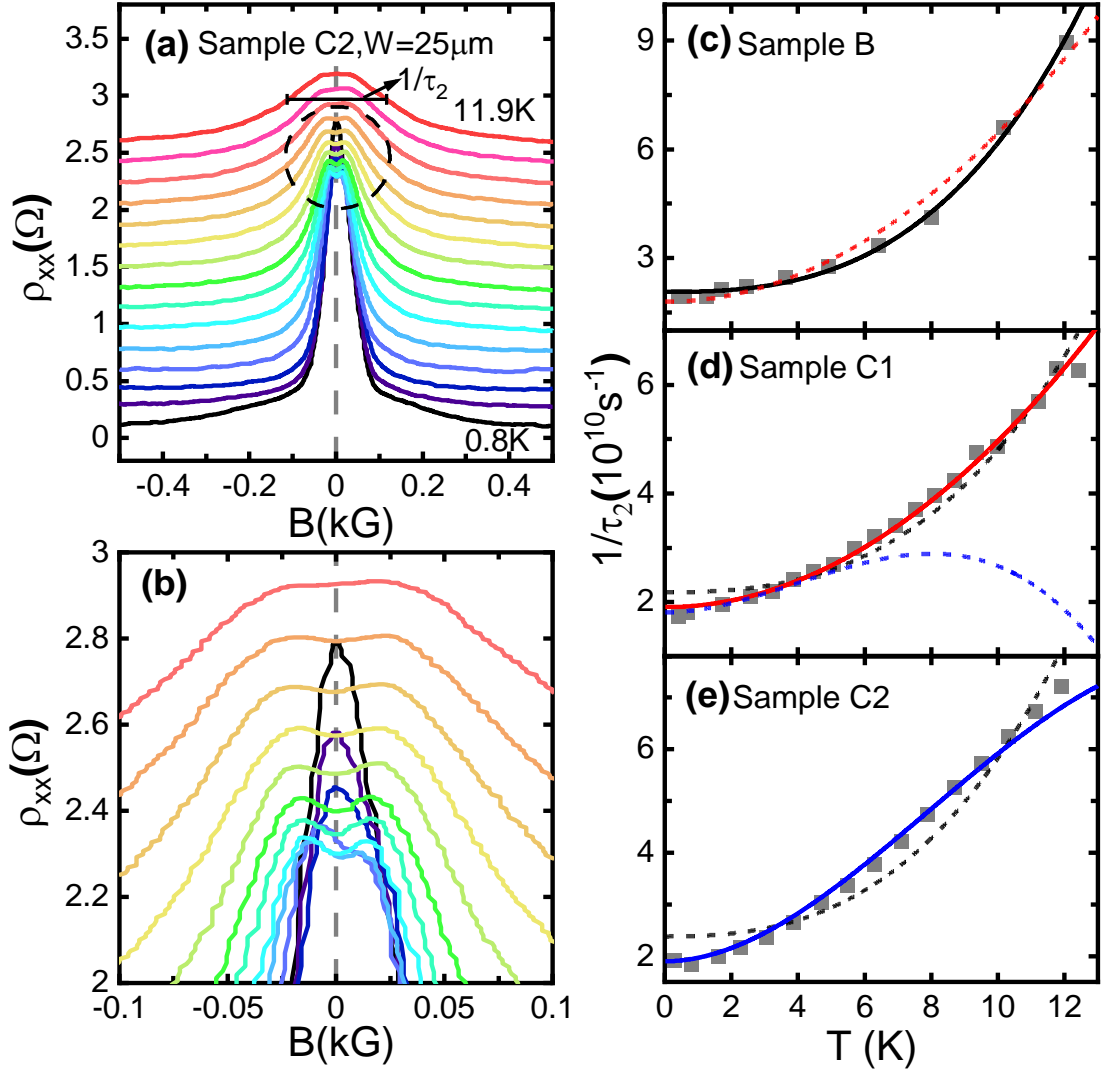

FIG. 3. (a) Magnetoresistivity of Sample C2 at different temperatures from 0.8 K to 11.9 K, with $1/\tau_2$ extracted from the full width half maximum of NNMR. (b) Zoom in of Fig. 3 (a) shows details of NNMR near zero field. Gray squares in (c)(d)(e) are the extracted data from 25 μm Hall bars of Sample B, C1, and C2 respectively. The data is fitted assuming $1/\tau_2(T) = 1/\tau_{2,ee}(T) + 1/\tau_{2,0}$, with three different functions of $1/\tau_{2,ee}$: Eq. (2) in black, Eq. (3) in blue, and $1/\tau_{2,ee} \propto T^2$ in red. (c) For Sample B with $r_s = 1.09$, Eq. (2) fits quite well. (d) For Sample C1 with $r_s = 0.92$, $1/\tau_{2,ee}$ is proportional to $T^2$. (e) For Sample C2 with $r_s = 0.86$, Eq. (3) fits well below 10 K. Difference among these samples shows density related crossover between viscous fluid and viscous gas.

The shape of BNMR (Fig. 1) is a parabolic curve[18]:

$$\rho_{xx}(B) = \rho_{xx}(0)(1 - \beta B^2), \qquad (4)$$

attributed to a field-independent correction to longitudinal magnetoconductance $\Delta\sigma_{xx}$ in previous work[35]. Here $\rho_{xx}(0)$ does not contain contribution from NNMR. Similar 'bell-shaped' NMR[36, 37] in lower mobility samples was explained through interaction correction theories[23, 35, 38] and $\Delta\sigma_{xx}$ was shown to be proportional to $-\ln(T)$ or $1/\sqrt{T}$ in different regimes. However, in recent reports[18-20], the existing theory is unable to explain BNMR in high purity samples for its peculiar relation between $\beta$ and $T$, and, more importantly, the influence of an in-plane magnetic field (see ref.18 and here in Fig. 2(a)). We confirm that BNMR is a size-dependent effect [19] since $\beta$ is approximately proportional to sample width $W$ (Inset of Fig. 2(b)) and this might explain why BNMR was not observed in extremely narrow samples ($W \sim 5\mu m$) [13]. Overall, although the $W$-dependence may hint on the viscosity origin, the nature of the BNMR is not clear at this point and its description remains at a phenomenological level.

*Microwave-induced resistance oscillations and 'second harmonics'.* -As shown in Fig. 1, the viscosity effect has a strong dependence of $W$, the sample width. Given that hydrodynamic theory together with memory effect may satisfactorily explain MIRO, as suggested by [29], this section will focus on $W$-dependence of MIRO. In our experiment, MW with a frequency of 102.4 GHz is produced by a Gunn oscillator and its power can be attenuated by a programmable rotary vane attenuator.

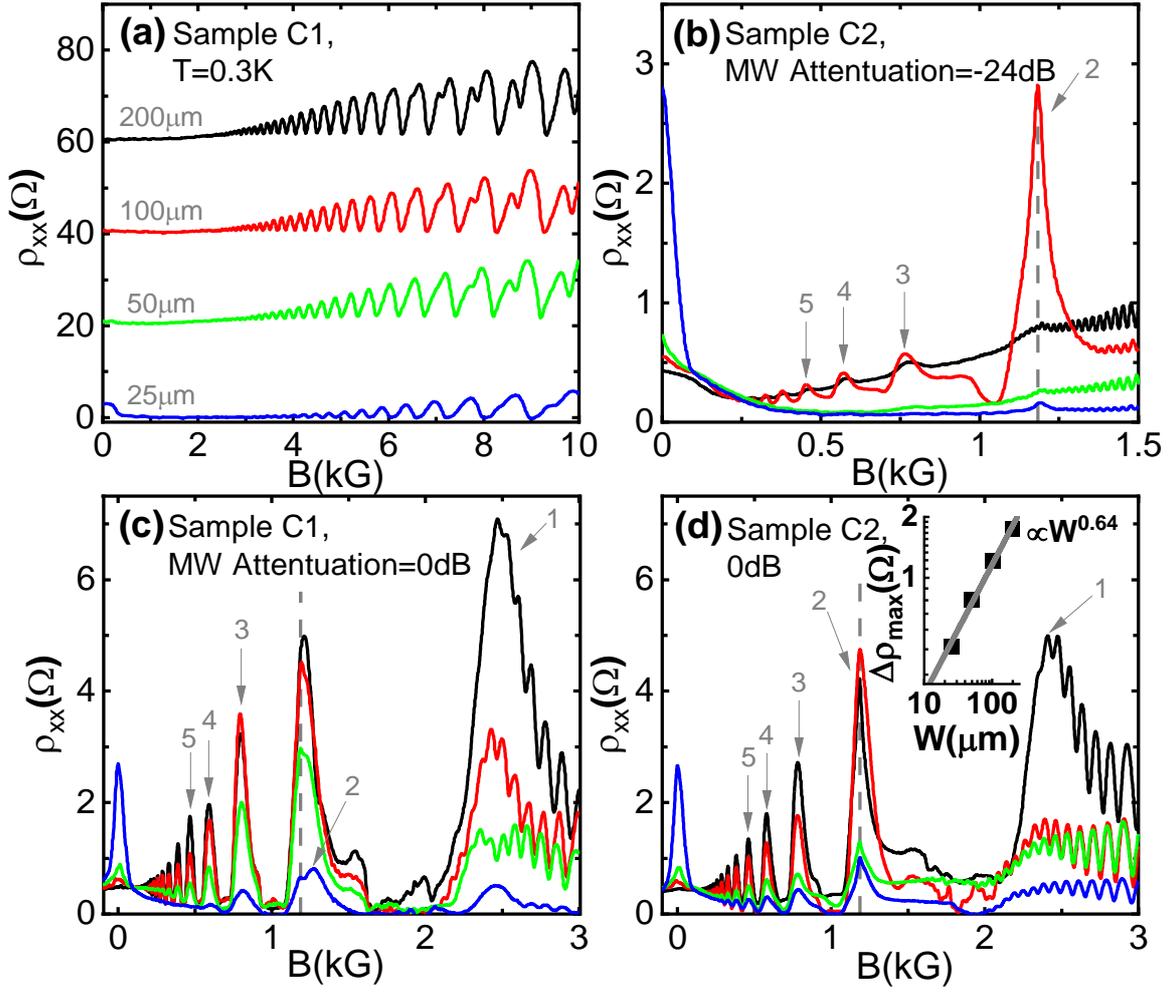

FIG. 4. a) SdH oscillations of Sample C1 without MW. For clarity, each $\rho_{xx}$ trace except that of 25 μm is consecutively shifted upwards by 20 Ω. Traces of PR at 0.3 K with different colors are taken from Hall bars with different widths: 200 μm (black), 100 μm (red), 50 μm (green), 25 μm (blue). Gray dashed line indicates the '2nd harmonic' peak at exactly $\omega = 2\omega_c$ and first five MIRO peaks are marked with gray arrows. (b) MIRO of Sample C2 with attenuation equal to -24dB shows a sharp '2nd harmonic' peak for 100 μm Hall bar. (c) and (d): MIRO of Sample C1 and C2 without MW attenuation. The inset of (d) shows the relation of MIRO amplitude and sample width when $\omega > 2\omega_c$. The amplitude is fitted with the field-dependent term removed, following the method of paper[39].

Fig. 4(c) and (d) demonstrate the $W$-dependence of the PR. For Hall bar with different $W$, Shubonikov-de Hass (SdH) oscillation is shown in Fig. 4(a) with almost invariant quantum scattering time and oscillation magnitude, especially for Hall bar widths equal to 200 μm, 100 μm and 50 μm, i.e. sample processing does not affect the quality of the 2DEG within this $W$ range. Compared with SdH oscillation, MIRO of all samples markedly depends on $W$ in an opposite way of NNMR. It's not surprising that stronger MIRO coexists with weaker NNMR from hydrodynamical point of view. In a Poiseuille flow, sample width determines the maximum of electron velocity, and therefore, there's a positive correlation between zero-field conductance/photoconductance and $W$. However, zero-field resistance is reciprocal of conductance, while PR is proportional to photoconductance because of large Hall resistance at finite magnetic field, giving rise to opposite $W$-dependence between NNMR and MIRO. Quantitively, here the amplitude of PR peak near $\omega = \omega_c$, *i.e.*, the first peak, is approximately linear with $W$, strongly favoring the theoretical prediction[29]. The amplitude $\Delta\rho_{\max}$ in lower field ($\omega > 2\omega_c$) is found to be proportional to $W^{0.64}$ (This approximate relation is extracted from the inset of Fig. 4 (d).), which differs from the linear dependence of $W$ for the wide samples, *i.e.*, $W \gg v_F^\eta \tau_2$.

A sharp peak located precisely at $\omega = 2\omega_c$ is observed in all the four samples, more clearly in relatively narrow Hall bars. Examples are shown in Fig. 4. In particular, strong '2nd harmonic' is observed in $100\ \mu m$ Hall bar of wafer C2 (Fig. 4(b)). There's always a tiny peak at $\omega = 2\omega_c$ when MIRO almost disappears in Fig. 4(b). This can be attributed to the fact that the amplitude of MIRO declines faster when the MW power is weakened. According to the viscosity theory [26, 29], we can understand the competition between MIRO and the '2nd harmonic' when $W$ is varied. On one hand, transverse magnetosonic waves, origin of '2nd harmonic', prevail only when viscoelastic resonance overweighs magnetoplasmon effect. Relevant upper limit for $W$ is the characteristic wavelength of magnetoplasmons. On the other hand, the dc PR peak originating from ac viscoelastic resonance at $\omega = 2\omega_c$ is size-dependent just like MIRO, *i.e.*, narrower sample corresponds to weaker PR. On balance there exists a optimal sample size for '2nd harmonic' peak and for Sample C, the value appears to be around $100\ \mu m$. Our data strongly support the viscoelastic nature of MIRO and the 'second harmonic' observed in ultrahigh-mobility 2DEG under microwave radiation.

*Conclusion.* -Width dependence of NMR in ultrahigh-mobility 2DEG is in accordance with recent viscoelastic dynamics theory. The previously reported two types of NMR are, for the first time, discriminated in a hydrodynamical framework. Measurements with different temperature show quantitatively that, there exists a crossover regime around $r_s \sim 1$ sharing some common properties between viscous liquid and viscous gas of electrons. Furthermore, existence of '2nd harmonic' peak and width dependence of MIRO also indicate that 2DEG in semi-classical regime (B < a few kG) should be taken as a viscous fluid. Certain caveats at quantitative level are noted, and should be addressed in future investigations. Overall, the intriguing experimental discoveries reported here strongly support viscous fluid theory of electrons in ultrahigh-mobility GaAs/AlGaAs QWs. Referring to Table I, we would like to mention that, as the GaAs/AlGaAs QWs have reached the ultrahigh-mobility regime, it becomes necessary to consider the hydrodynamic effect in assessing the quality of the 2DEG, in addition to the Drude mobility from impurity scatterings.

*Acknowledgements.* The work at PKU was funded by the National Key R&D Program of China (Grants No. 2017YFA0303300 and 2019YFA0308400), by the Strategic Priority Research Program of Chinese Academy of Sciences (Grant No. XDB28000000). The work at Princeton was funded by the Gordon and Betty Moore Foundation through the EPiQS initiative Grant No. GBMF4420, by the National Science Foundation MRSEC Grant No. DMR-1420541.